\documentclass[traditabstract]{aa} % for the abstract without structuration 
\usepackage{graphicx}
\usepackage{txfonts}
\begin{document}
   \title{The spectroscopic evolution of the recurrent nova T Pyxidis
during its 2011 outburst}
   \subtitle{I. The optically thick phase and the origin of moving
lines in novae}

   \author{S. N. Shore\inst{1}, T. Augusteijn\inst{2}, A. Ederoclite
\inst{3,4}and H. Uthas\inst{5}}
        \institute{
 Dipartimento di Fisica ``Enrico Fermi'', Universit\`a di Pisa, and INFN- Sezione Pisa, largo
B. Pontecorvo 3, I-56127 Pisa, Italy; \email{shore@df.unipi.it}
  \and
Nordic Optical Telescope, Apartado 474, E-38700 Santa Cruz de La Palma,
Santa Cruz de Tenerife,Spain; \email{tau@not.iac.es}
\and
Instituto de Astrof\'isica de Canarias, E-38200, La Laguna, Tenerife, Spain; \email{ale@iac.es}
\and 
 Departamento de Astrof\'isica, Universidad de La Laguna, E-38206 La Laguna, Tenerife, Spain
 \and
 Department of Astronomy, Columbia University, 550 W 120th Street, New
York, NY 10027, USA; \email{helena@astro.columbia.edu}
  }
     
              \date{Received ---; accepted ---}

 \abstract{We aim to derive the physical properties of the recurrent nova T Pyx and  the  structure of the ejecta  during the early stages of expansion of the 2011 outburst.
 The nova was observed with high resolution spectroscopy ($R \approx 65000$ spectroscopy, beginning 1 day after discovery of the outburst and continuing through the last visibility of the star at the end of May 2011.  The interstellar absorption lines of Na I, Ca II, CH, CH$^+$, and archival H I 21 cm emission line observations have been used to determine a kinematic distance.  Interstellar diffuse absorption features have been used to determine the extinction independent of previous assumptions.   Sample Fe-peak line profiles show the optical depth and radial velocity evolution of the discrete components. 
We propose a distance to T Pyx $\geq$4.5kpc, with a strict lower limit of 3.5 kpc (the previously accepted distance). We derive an extinction, E(B-V)$\approx$0.5$\pm$0.1, that is higher than previous estimates.  The first observation, Apr. 15, displayed He I, He II, C III, and N III emission lines and a maximum velocity on P Cyg profiles of the Balmer and He I lines of $\approx$2500 km s$^{-1}$ characteristic of the fireball stage.  These ions were undetectable in the second spectrum, Apr. 23, and we use the recombination time to estimate the mass of the ejecta, $10^{-5}f$M$_\odot$ for a filling factor $f$.  Numerous absorption line systems were detected on the Balmer, Fe-peak, Ca II, and Na I  lines, mirrored in broader emission line components,  that showed an ``accelerated'' displacement in velocity.  We also show that the time sequence of these absorptions, which are common to all lines and arise only in the ejecta,  can be described by  recombination front moving outward in the expanding gas without either a stellar wind or circumstellar collisions.  By the end of May, the ejecta were showing signs of turning optically thin in the ultraviolet. }

   \keywords{Stars-individual(T Pyx), physical processes, novae }

  \titlerunning{The 2011 outburst of T Pyx. I} \authorrunning{S. N.
Shore et al.}

  \maketitle

\section{Introduction}

Classical novae (CNe) are thought to be the result of the thermonuclear runaway on the surface of a white dwarf {\bf that} is accreting mass from a Roche-lobe filling companion (see the recent compilation by Bode \& Evans 2008). Since the system survives the eruption, a CN is a repeating  phenomenon although the Òrecurrence timeÓ (i.e. the time between two consecutive explosions) is $>$10$^3$ yrs. Recurrent novae (RNe) display repeat  outbursts  with a much higher frequency, on intervals of $<$100 yrs,  T Pyx bears greater resemblence to the CNe in having a Roche-lobe filling companion and orbital period shorter than 1 day. It was discovered in 1890 and underwent outbursts at nearly regular intervals in 1902, 1920, 1944, and 1967.  It then went into a dormant state for unknown reasons.  
The wait ended when the sixth  outburst was discovered on the 2011 Apr. 14.29 (JD 2455665.79, IAUC 9205).  The orbital period has been determined by Uthas et al. (2010) to be 1.83~h by spectroscopy during the hiatus interval.   They derive a WD mass of $M_{WD}=0.7\pm 0.2 M_\odot$.  This depends on the assumed mass for the companion, M$_{\rm comp}$ and is, in fact, a lower limit since their radial velocity curve only yields a robust estimate for the mass {\it ratio}, $M_{\rm WD}/M_{\rm comp} \approx 0.2$.  In contrast, model for RNe need a high-mass WD to have such high recurrence rates (see Starrfield et al. 1985, Starrfield \& Iliadis 2008). 
%\footnote{The long quiescence has produced a variety of hypothesis on the ``real'' nature of this object, e.g. Schaefer, Pagnotta, \& Shara (2010) even %suggested that T Pyx is not a genuine RN but rather a CN whose explosion occurred in 1866 and now showing dwarf nova outburst.}  
\section{Observational Data}
%  \small{
\begin{center}
Table 1: Journal of NOT observations \\
\begin{tabular}{ccc}
\hline
Date (2011) & Time (UT) & JD 2400000+ \\
\hline
Apr. 15 & 22:07 & 55667.42 \\
Apr. 23 & 22:51 & 55675.46 \\
May 6  & 20:56 & 55688.38 \\ 
May 8 & 20:55 & 55690.37 \\
 May 15 & 21:04 & 55697.38 \\ 
May 21 & 20:47 & 55703.37 \\ 
May 30  & 21:00 & 55712.38 \\
\hline
\end{tabular}
\end{center}
Our data set consists of spectra taken between
2011 Apr 15 and 2011 May 30 with the 2.6 m Nordic Optical Telescope
(NOT) fiber-optic echelle spectrograph (FIES, program P40-403) with a dispersion of 0.023\AA px$^{-1}$ in
high-resolution mode covering the spectral interval from 3635 to
7364\AA and 0.035\AA px$^{-1}$ in medium-resolution mode covering
the spectral interval from 3680 to 7300\AA\ (see Table 1).  Exposures ranged from 60
s to 900 s. The sequence was not absolutely calibrated. All NOT
spectra were reduced using IRAF, FIESTools, and IDL.
\footnote{IRAF is
  distributed by the National Optical Astronomy Observatories, which
  are operated by the Association of Universities for Research in
  Astronomy, Inc., under cooperative agreement with the US National
  Science Foundation.} 
  %}
%\section{Analysis}

\section{Distance and extinction estimates}

The distance and extinction to T Pyx are the parameters with the greatest uncertainty.  Selvelli et al  (2008) and Schaefer (2010) provide the most recent compilations of the estimates for T Pyx, giving a distance of 3.5$\pm1$ kpc and E(B-V)=0.25$\pm$0.02 and for purposes of this Letter we refer the reader to Tables 26 and 29 of Schaefer (2010) for details.  Although some spectroscopy during the extended minimum has been included, with inconclusive results, almost all estimates of both distance and reddening have been based on photometry.   We have attempted to obtain limits on both spectroscopically.   The NOT spectra contain a  number of atomic and molecular absorption lines, all of which are resolved into multiple components or display asymmetric profiles in a manner that indicates more than one contributor.  These are listed in table 2 and displayed in Fig. 1.   Neutral hydrogen 21 cm line profiles are available for this Galactic position ($l_{II}$=257.2, $b_{II}$=+9.6) from the LAB survey (Kalberla et al.  2005) with a broad component extending from +32 to +52 km s$^{-1}$ that spans the radial velocity interval of the atomic and molecular absorption lines.  
Using the Galactic rotation curve for the 3rd quadrant (Brand \& Blitz 1993) we find a {\it minimum} distance consistent with previous estimates, 4.0 kpc, for $\Theta_0$=220 km s$^{-1}$ based on v$_{\rm rad}$ for CH 4300\AA.   The center of mass velocity of the binary is still uncertain and may be variable (Uthas et al. 2010) but there may be another way to obtain it.  In the JD 55712 spectrum,  the emission lines (e.g. Fe II RMT 42, H Balmer) show two nearly symmetric emission  peaks, and the same holds for the weak emergent  [O I] 5577, 6300\AA\ emission with a separation of about 700 km s$^{-1}$ and mean velocity of +60$\pm$3 km s$^{-1}$.\footnote{In their orbital solution of T Pyx using the He II 4686\AA\ emission line during quiescence, Uthas et al. (2010) were unable to derive a definitive systematic velocity for the nova but all of their solutions were at more negative velocity than +20 km s$^{-1}$.
We suggest that there may be a residual uncertainty due to the profile of the He II line and/or shell emission that compromises the center of mass velocity but not the amplitude of the variations, 17.9$\pm$1.6 km s$^{-1}$.}  If real, this is corresponds to a distance of $\approx$5 kpc.   There are no independent interstellar measurements from nearby OB stars in the archives.  We therefore conclude that the distance to T Pyx is  $>$4 and $\leq$5  kpc.
        
    %\small{
\begin{center}
Table 2: Heliocentric interstellar line velocities \\
\begin{tabular}{cccc}
\hline
Identification & $\lambda$ (\AA) & abs/em & v$_{\rm rad}$ km s$^{-1}$\\
\hline
Ca II K & 3933.66 & abs &  34.5, 42.3\\
CH$^+$ & 4232.54 & abs & 38.4 \\
CH & 4300.31 & abs & 49.5 \\
Na I D1 & 5889.95 & abs & 30.7, 40.8 \\
Na I D2 & 5895.92 & abs & 30.4, 42.5 \\
H I & 21 cm & em & 11.4, 21.9, 34.8, 81.1 \\
\hline
\end{tabular}
\end{center}
%}
    
The diffuse interstellar bands (DIBs), taken from the list by Friedman et al. (2010) for those best correlated with extinction, provide an independent estimate of E(B-V) for T Pyx.  The equivalent widths are given in table 3, the uncertainty is $\pm$10\% for the measurements.  Combining  5780, 6205, 6196, and 6283 gives E(B-V) = 0.49$\pm$0.17.  The strongest DIBs, 5780.5\AA\ and 6613.7\AA\, are shown in Fig. 1 along with the interstellar absorption lines listed in Table 2.    The effect of this revision on the luminosity, both in quiescence and outburst is obvious but the larger extinction also
affects the derived properties of the WD based on the continuum from 1200-3000\AA\ (Selvelli et al. 2008). The increase in
E(B-V) increases the UV spectral gradient with wavelength such that
the effective temperature of the WD is lower than earlier
determinations (but how much this changes depends on a model atmosphere
analysis that we will discuss in the next paper in this series). 

    %\small{
\begin{center}
Table 3: Diffuse Interstellar Bands \\
\begin{tabular}{cc|cc|cc}
\hline
$\lambda$ (\AA) & EW (m\AA) & $\lambda$  & EW  & $\lambda$ & EW \\
\hline
5780 & 280 & 6204 & 104 & 6613 & 49 \\
5799 & 63 &  6283 & 284 & 6623 & 62 \\
6196 & 30 & 6379 & 28 &  6532 & 22 \\
6203 & 65: & 5632 & 22 &  7224 & 50 \\
\hline
\end{tabular}
\end{center}
%}
 \begin{figure}
   \centering
   \includegraphics[width=9cm]{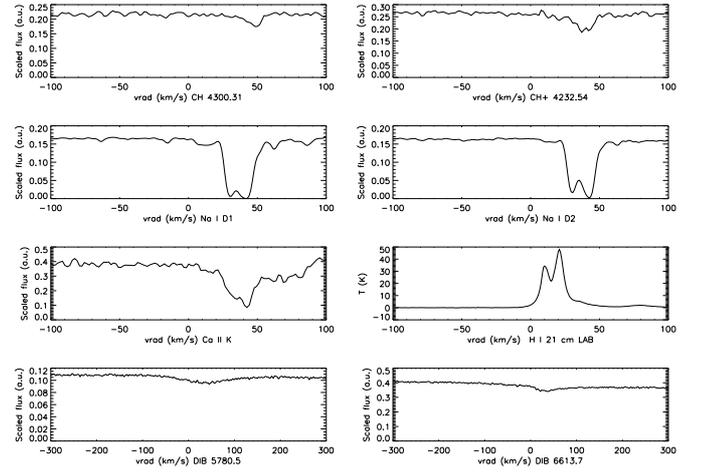}
   \caption{Interstellar absorption lines from the spectrum taken on JD 55697.  The radial velocities are heliocentric (conversion for the H I 21 cm LAB spectrum, using 220 km s$^{-1}$ for the LSR)  (see text for discussion).  Note the change in velocity scale for the DIBs in the lowest two panels.}
    \end{figure}

\section{Discrete absorption features}
    
Our first observation, JD 55667, was one day after the discovery and showed  a fireball spectrum (see, e.g.Schwarz et al. 2001) dominated by P Cygni profiles of the H I Balmer lines (Figs. 2 and B.1), He I, and and He II, for which the absorption extended without a terminal edge to $\approx$2500 km s$^{-1}$; {\bf Fig. 4  shows an example of the fireball stage profile of He I 5875\AA\ profile compared to the later appearing Na I D}.   The emission complex at 4640\AA\, due to C III and N III, was well resolved into the two main contributors without an absorption trough on either.  Subsequent spectra showed only low ionization Fe-peak transitions, Ca II H and K, and the Na I D doublet, all of which displayed increasingly complex sets of discrete high negative radial velocity absorption components (DACs hereafter).  
 \begin{figure}
   \centering
   \includegraphics[width=9cm,height=4cm]{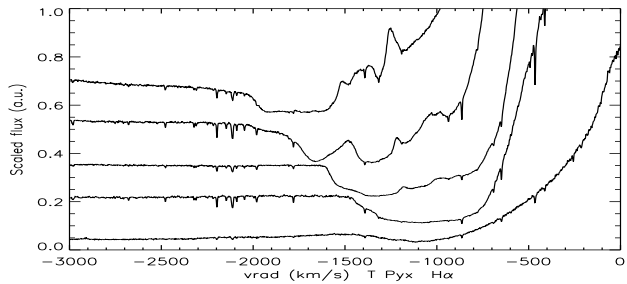}
   \caption{H$\alpha$ profile variations showing the development of the DACs.  From bottom to top, JD 55675, 55688, 55697, 55703, and  55712.  Only the negative radial velocity components are shown, fluxes are scaled but the continuum is otherwise unnormalized. }
    \end{figure}
The velocity components identified on the Fe II and Balmer lines on JD
55703 are at -2964,-1661,-1543,-1433,-1303,-1153,-1056,-952, -799 km
s$^{-1}$ with mean separation $\Delta v_{rad} =122\pm21$ km s$^{-1}$
and $\Delta v \leq 50$ km s$^{-1}$ for every component based on the Fe II RMT 42 lines\footnote{RMT refers to the {\it Revised Multiplet Tables}, Moore 1945.}; the minimum width was $\approx 5$ km s$^{-1}$.  Subsequent velocity shifts are consistent with a linear velocity law for the ejecta.  The broader {\it features} are likely blends of still optically thick
individual components. If the width indicates the filament linear thickness  then their  filling factor  may be quite low but their individual optical depths high (see below) .   The complexity of the radiative coupling between various levels of the
iron group contributors to the atomic opacity is especially well
illustrated by the behavior of three multiplets, Fe II RMT 27, 28, and
42. Of the first two, Fe II 4173.461, 4178.862, and 4233.167 showed a
moving feature at -1200 km s$^{-1}$ that displaced systematically
toward higher velocity, approximately coincident with the same feature
in Fe II RMT 42 and the Balmer lines.  These strong absorptions were short-lived.  A third line in the multiplet, 5607.14\AA, showed
nothing.  All three members of RMT 42, in contrast, showed identical evolution (we display only 5018.44\AA, the strongest member of the triplet) and showed persistent narrow lines even in the last observation.  The lower level of RMT 26 and 27 is fed by transitions at 2500-2600\AA\ that, according to the {\it Swift} UVOT observations were becoming transparent by JD 55703 (the analysis of the ultraviolet variations is postponed to a later paper).  {\bf Examples of the line profiles for Fe II are shown in Fig. 3.}

The fine structure that developed on the absorption lines was also
mirrored in the emission component but with less complexity. This
would be expected for a shell in expansion, the total optical depth
decreases with a similarity scaling as $\tau \sim t^{-2}$ with
individual features changing depending on the rates of ionization and
recombination. In this interpretation, the fine structure was present
from the earliest epoch, likely at the moment of ejection, but the
visibility -- contrast -- of the features changed over time. The
minimum velocity at which the fine structure absorption features
appeared on the Fe II RMT42 lines was -572$\pm$5 km s$^{-1}$ (JD
55688), in the previous spectrum these were blended within the broad
absorption trough whose minimum was at -635 km s$^{-1}$. However, the
P Cyg features were separated from the emission (see Fig. 3) with a
minimum at -270 km s$^{-1}$. The strongest line of the multiplet,
5169\AA, was the first to display the symmetric peaks and
systematically displayed the strongest narrow absorption components.
Note that the emission wing extended to the same maximum velocity as
the absorption throughout this sequence.  The Fe II lines  on  JD 55712, are consistent with the
development of a ``thin shell'' structure, the symmetric peaks at
-376 and +476 km s$^{-1}$. The central velocity, $v_{\rm rad} = +60$
km s$^{-1}$, is {\it consistent} with the detection of a high positive
velocity component on the interstellar lines at $>$50
km s$^{-1}$ and with a distance $\approx$ 5 kpc if this is the radial
velocity of the T Pyx system. The emission peak shifted
systematically from +20 km s$^{-1}$ to +60 km s$^{-1}$ before JD 55712.
 \begin{figure}
   \centering
   \includegraphics[width=9.2cm]{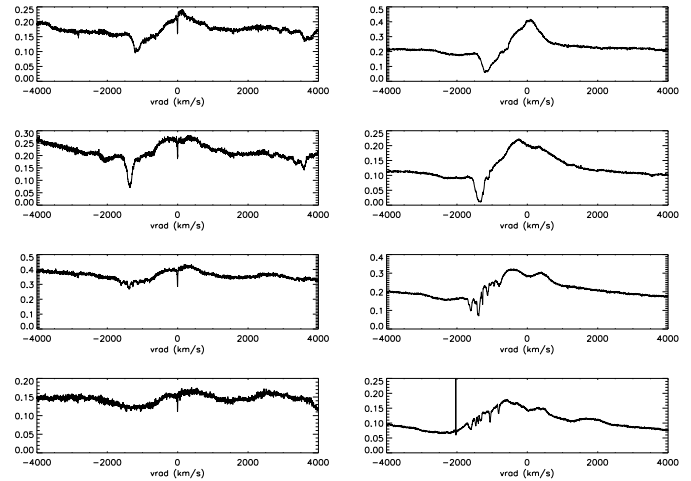}
   \caption{Profile development for Fe II 4233.14\AA\ (RMT 26, left) and Fe II 5018.44\AA\ (RMT 42, right) .  From top to bottom, the dates are JD 55688, 55697,  55703, and 55712.   In addition to the absorption components, note the change sin the emission near +2000 km s$^{-1}$ (see text for discussion).}
    \end{figure}    

The Na I D lines, the only strong resonance transitions of a neutral species, are especially important for understanding the profile variations.   They showed the broad features but  as the optical depth decreased, these too resolved into an ensemble of narrow absorption components; see Fig. 4.   The broad D1 feature, which shifted -1900 to -2200 km s$^{-1}$, had no D2 counterpart.  For every other, narrower, D1 component there is a corresponding feature from D2.  Chance coincidences, overlaps between some D1 feature with a different one from D2,  account for their intensities,     Discrete components as narrow as 10 km s$^{-1}$ are detected and suggests that {\it all} of the broader features are, in fact, composite.  Their individual optical depths decreased in time.    As the outburst progressed, the emission component also broadened toward higher positive velocity with broad peaks that mirrored the absorption line velocities but with less distinct separation.  This can be interpreted within the scenario of a ballistically expanding shell as a recombination front progressing through the ejecta while the overall optical depth of the expanding ejecta decreased with time.  If this is correct, the knots were present even in the earliest stage of the expansion but hidden by the more opaque overlying layers.  The same effect was seen on the Fe II and H I Balmer lines but, arising from excited states, the interpretation is not so straightforward.  The narrowest components, which agree between the different ions, were only 10 km s$^{-1}$ wide, indicating a low spatial filling factor but high individual column densities.
 \begin{figure}
   \centering
   \includegraphics[width=9cm,height=4cm]{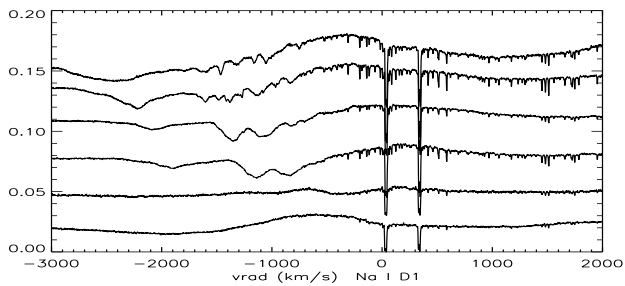}
   \caption{{\bf The sequence of Na I D line variations for JD 55667, 55675, 55688, 55697,  55703, and 55712  (from bottom to top).  Only He I 5875\AA\ is present in the first (JD 55667, bottom) spectrum.}  The progressive displacement of the DACs, with the doublet .  The spectrum from JD 55690 is not included because of the lower S/N ratio (see text) }
    \end{figure}
    
\section{Discussion: The mass and structure of the ejecta}

Perhaps the most significant result from these observations is the insight they provide the origin of the moving line systems observed in nova ejecta. It is unusual to have so long a plateau for a recurrent nova (although not for a classical nova) in the visible during which the UV is optically thick in lines from neutrals and singly ionized species.  We do not know, at this point, {\it why} the ejecta from T Pyx are so massive.  This episode is not unusual in rise, duration, or amplitude compared with previous outbursts and suggests that the ejection is not unusually massive relative to the historical events despite the long hiatus.  Thus,  the details provided by the NOT spectra become especially significant: this is showing a generic behavior that is also seen in  classical novae (Payne-Gaposchkin 1957, McLaughlin 1954, 1964).
 
The recombination of He II following the first observation, an interval of about one week, provides an estimate of the electron density, $n_e \approx 2\times 10^6$ cm$^{-3}$ for the outer part of the ejecta at -2500 km s$^{-1}$ at a radius of $1.5\times 10^{15}$ cm.  Assuming a ballistic expansion with an $r^{-3}$ radial dependence and an innermost of 500 km s$^{-1}$ based on the separation of the inner emission peaks on the Fe II profiles, the mass of a spherical, filled ejecta is $\approx 2\times 10^{-5}f$M$_\odot$, where $f$ is the radially constant filling factor.  This is consistent with the long duration of the Fe-curtain phase, more than 50 days and of the same order as in classical novae for which such masses are normal but must be an upper limit.  If, however, the shell is severely fragmented, as appears the case from the DACs, but every line of sight encounters at least one such optically thick filament, then the mass {\it could} be lower by a large factor.  A filling factor $f \sim 0.01$ would reduce the mass estimate to that normally found for recurrent novae  (e.g. Anupama 2009, Schaefer 2010) but requires that the individual filaments or substructures are extremely opaque and completely covering.  A test is whether there is a constant bolometric luminosity phase such as that found for classical novae (see e.g. Shore 2008){\bf ,} but that requires continued ultraviolet observations that are not yet completed.
  
 \begin{figure}
   \centering
   \includegraphics[width=9cm,height=4cm]{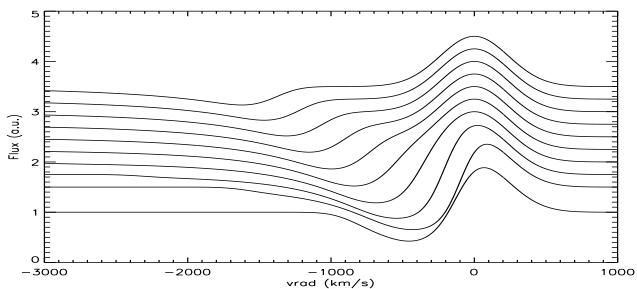}
   \caption{Schematic model for the line variations, assuming a recombination front moving outward within separated ejecta with a constant mass and linear velocity law.  The emission line profile is kept arbitrarily constant throughout the interval see text)}
    \end{figure}
  
 As a proof-of-concept, we show in Fig. 5 a schematic  model for the absorption line evolution that applies most simply to the Na I resonance lines (Fig. 4).  This assumed a recombination front moving outward in the expanding ejecta.  A constant  velocity gradient with an $r^{-3}$ density variation were  imposed and the recombination was integrated through the ejecta.  Each profile represents one time step, the timescale being the recombination time from the disappearance of C III/N III 4640\AA\ and the He II lines between the first and second NOT observation, about 5 days.  The intrinsic line profile was assumed to be gaussian, but the details are not important for this simulation.  The decrease in the total optical depth, scaling as $t^{-2}$, and the separation of the absorption component are reproduced.  The individual fine structure features are not included, they would be part of the absorption in any velocity interval, nor is the emission (only an invariant gaussian was used for visualization of the absorption since the ejecta are assumed to be detached from the white dwarf.   The line evolution can thus be understood assuming only that the structure was frozen in the ejecta at the time of expulsion.  There is no need for either colliding shocks and winds (Kato \& Hachisu 2007) or subsequent collisions with circum-system material (e.g. Williams et al. 2008, Williams \& Mason 2010), although the latter may also occur4.  The evolution of the different DACs can be explained the changes in the optical depth of specific transitions in the ultraviolet as the ejecta expand.  Finally, the last NOT spectrum shows weak emission appearing on [O I] 5577 and 6300\AA\ with a double peaked structure that is consistent with  with the innermost part of the ejecta becoming optically thin in the ultraviolet and beginning to again display both He I and [O I] emission lines.

        \begin{acknowledgements}

We warmly thank N. Morrison, Univ. of Toledo - Ritter Observatory for providing the archival echelle spectra of V705 Cas, P. Kalberla for correspondence regarding the LAB 21 cm line profiles,  and G. Schwarz, S. Starrfield, G. M. Wahlgren, and F. Walter for valuable discussions.    AE acknowledges support by the Spanish {\it Plan Nacional de Astronom\'ia y Astrof\'isica} under the grant AYA2008-06311-CO2-01.  We have made extensive use of the Astrophysics Data System (ADS), SIMBAD (CDS), and the MAST archive (STScI) during this work.  We thank the referee, R. Gehrz, for comments that clarified the text.
 
\end{acknowledgements}

\appendix
\section{V705 Cyg 1993 and discrete absorption systems on Na I D}

The D lines have been noted to display P Cyg profiles occasionally in the literature, notably during the first phase of the DQ Her 1934 outburst (see McLaughlin 1954, 1964) but based on photographic, uncalibrated, and relatively low resolution spectra this might be explained as variations in the complex of emission and absorption lines normally seen in the initial stage of the expansion.  The first classical  CO nova to show this phenomenon unambiguously after the advent of CCD detectors,  V705 Cas 1993, is shown in Fig. A.1.  This sequence shows the optically thick phase, during the first two weeks of the outburst.  The line pairs shifted systematically to higher velocity  with an increasing optical depth, indicated by the increase in the F(5895)/F(5889) ratio, with FWHM$\approx$300 km s$^{-1}$.  The sequence ended quite early in the outburst, the nova formed dust about two months later (Shore et al. 1994), as did DQ Her and several other novae in which the Na I D multiple components and moving DACs were historically observed.   The NOT sequence for T Pyx covered a longer interval.   Similar behavior has recently been detailed by Sadakane et al. (2010) for V1280 Sco.
    
 \begin{figure}
   \centering
   \includegraphics[width=9cm,height=5cm]{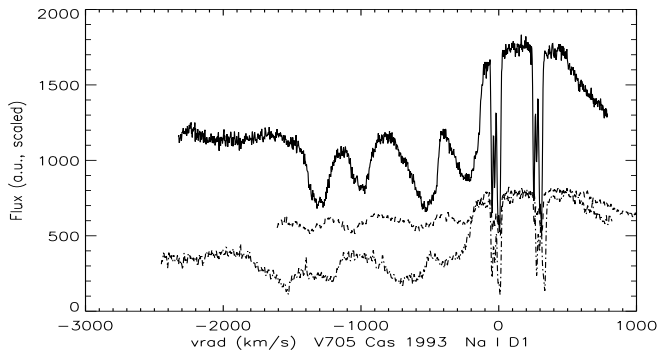}
   \caption{V705 Cas (Ritter Observatory spectra of the optically thick stage of the CO nova Nova Cas 1993, Na I D line variations; top: 1993 Dec. 27; middle: 1993 Dec. 29, 1994; bottom: 1994 Jan. 5 (see text for details)}
    \end{figure}
    
    \section{ONLINE FIGURES}
    
 \begin{figure}
   \centering
   \includegraphics[width=9cm,height=4cm]{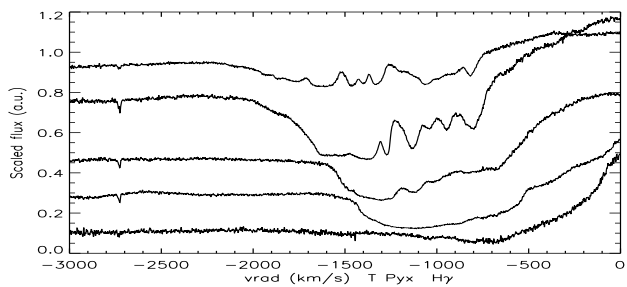}
   \caption{H$\gamma$ profile variations, as in Fig. 2}
    \end{figure}
    
\end{document}